# A Novel Approach for Cardiac Disease Prediction and Classification Using Intelligent Agents

Murugesan Kuttikrishnan
Department of Computer Science and Engineering
Anna University, Chennai, India
murugesan_k3@yahoo.com

Manjula Dhanabalachandran
Department of Computer Science and Engineering
Anna University, Chennai, India
manju@annauniv.edu

*Abstract*— The goal is to develop a novel approach for cardiac disease prediction and diagnosis using intelligent agents. Initially the symptoms are preprocessed using filter and wrapper based agents. The filter removes the missing or irrelevant symptoms. Wrapper is used to extract the data in the data set according to the threshold limits. Dependency of each symptom is identified using dependency checker agent. The classification is based on the prior and posterior probability of the symptoms with the evidence value. Finally the symptoms are classified in to five classes namely absence, starting, mild, moderate and serious. Using the cooperative approach the cardiac problem is solved and verified.

## I. INTRODUCTION

Intelligent agents are a new paradigm for developing software applications. More than this, agent-based computing has been hailed as 'the next significant breakthrough in software development' (Sargent, 1992), and 'the new revolution in software' (Ovum, 1994). Currently, agents are the focus of intense interest on the part of many sub-fields of computer science and artificial intelligence. Agents are being used in an increasingly wide variety of applications, ranging from comparatively small systems such as email filters to large, open, complex, mission critical systems such as air traffic control. At first sight, it may appear that such extremely different types of system can have little in common. And yet this is not the case: in both, the key abstraction used is that of an agent.

First, an agent is a computer system situated in some environment, and that is capable of autonomous action in this environment in order to meet its design objectives. Autonomy is a difficult concept to pin down precisely, but we mean it simply in the sense • that the system should be able to act without the direct intervention of humans (or other agents), and should have control over its own actions and internal state. It may be helpful to draw an analogy between the notion • of autonomy with respect to agents and encapsulation • with respect to object oriented systems. An object encapsulates some state, and has some control over this state in that it can only be accessed or modified via the methods that the object provides. Agents encapsulate state in just the same way. However, we also think of agents as encapsulating behavior, in addition to state. An object does not encapsulate behavior: it has no control over the execution of methods – if an object x invokes a method m on an object y, then y has no control over whether m is executed or not – it just is. In this sense, object y is not autonomous, as it has no control over its own actions. In contrast, we think of an agent as having exactly this kind of control over what actions it performs. Because of this distinction, we do not think of agents as invoking methods (actions) on agents – rather, we tend to think of them requesting actions to be performed. The decision about whether to act upon the request lies with the recipient. .

An intelligent agent is a computer system that is capable of flexible autonomous action in order to meet its design objectives. By flexible, we mean that the system must be:

- responsive: agents should perceive their environment (which may be the physical world, a user, a collection of agents, the Internet, etc.) and respond in a timely fashion to changes that occur in it.
- proactive: agents should not simply act in response to their environment, they should be able to exhibit opportunistic, goal-directed behavior and take the initiative where appropriate, and Applications of Intelligent Agents
- social: agents should be able to interact, when they deem appropriate, with other artificial agents and humans in order to complete their own problem solving and to help others with their activities.

Hereafter, when we use the term 'agent', it should be understood that we are using it as an abbreviation for 'intelligent agent'. Other researchers emphasize different aspects of agency (including, for example, mobility or adaptability). Naturally, some agents may have additional characteristics, and for certain types of applications, some attributes will be more important than others. However, we believe that it is the presence of all four attributes in a single software entity that provides the power of the agent paradigm and which distinguishes agent systems from related software paradigms – such as object-oriented systems, distributed sysems, and expert systems (see Wooldridge (1997) for a more detailed discussion). By an agent-based system, we mean one in which the key abstraction used is that of an agent. In principle, an agent-based system might be conceptualized in terms of agents, but implemented without any software structures corresponding to agents at all. We can again draw a parallel with object-oriented software, where it is entirely possible to design a system in terms of objects, but to implement it without the use of an object-oriented software environment. But this would at best be unusual, and at worst, counterproductive. A similar situation exists with agent technology; we therefore expect an agent-based system to be both designed and implemented in terms of agents. A number of software tools exist that allow a user to implement software systems as agents, and as societies of cooperating agents. Note that an agent-based system may contain any non-zero number of agents. The multi-agent case – where a system is designed and implemented as several interacting agents, is both more general and significantly more complex than the single-agent case. However, there are a number of situations where the single-agent case is appropriate.

## II. RELATED WORK

Traditional diagnosis in TCM requires long experiences and a high level of skill, and is subjective and deficient in quantitative diagnostic criteria. This seriously affects the reliability and repeatability of diagnosis and limits the popularization of TCM. So the focal problem that needs to be solved urgently is to construct methods or models to quantify the diagnosis in TCM.[1] Recently, a few researchers developed some methods and systems to modernize TCM. But most of them are built incorporating totally or partially rulebased reasoning model, which are lack of the feasibility of implementing all possible inference by chaining rules and limits their practical applications in clinical medicines.An attraction tool for managing various forms of uncertainty is Bayesian networks (BNs) [2], [3] which is able to represent knowledge with uncertainty and efficiently performing reasoning tasks.Naive Bayesian classifier (NBC) is a simplified form of BNs that assumes independence of the observations. Some research results [4], [5], [6] have demonstrated that the predictive performance of NBC can be competitive with more complicated classifiers.In this study, a novel computerized diagnostic model based on naive Bayesian classifier (NBC) is proposed. Firstly, a Bayesian network structure is learned from a database of cases [7] to find the symptom set that are dependent on the disease directly. Secondly, the symptom set is utilized as attributes of NBC and the mapping relationships between the symptom set and the disease are constructed. To reduce the dimensionality and improve the prediction accuracy of diagnostic model, symptom selection is requisite. Many feature selection methods, such as filters [8] and wrappers [9], have developed. But the dependency relationships among symptoms and the mapping relationships between symptom and syndrome are not considered in these methods, which are important to diagnosis in TCM.

To lower the influences of irrelative symptoms, the mutual information between each symptom and disease is computed based on information entropy theory [10], which is utilized to assess the significance of symptoms. The paper [11] presents a multiagent system for supporting physicians in performing clinical studies in real time. The multiagent system is specialized in the controlling of patients with respect to their appointment behavior. Novel types of agents are designed to play a special role as representatives for humans in the environment of clinical studies. OnkoNet mobile agents have been used successfully for patient-centric medical problems solving [12]., emerged from a project covering all relevant issues, from empirical process studies in cancer diagnosis/therapy, down to system implementation and validation. In the paper [13], a medical diagnosis multiagent system that is organized according to the principles of swarm intelligence is proposed. It consists of a large number of agents that interact with each other by simple indirect communication.

In the paper [14], a system called Feline composed of five autonomous agents (expert systems with some proprieties of the agents) endowed with medical knowledge is proposed. These agents cooperate to identify the causes of anemia at cats. The paper [39], also presents a development methodology for cooperating expert systems. In the paper [15], a Web-centric extension to a previously developed expert system specialized in the glaucoma diagnosis is proposed. The proposed telehealth solution publishes services of the developed Glaucoma Expert System on the World Wide Web. Each agent member of the CMDS system has problems solving capability and capacity (the notions are defined in [16, 17]). The capacity of an agent Agf (Agf U MDUAS) consists in the amount of problems that can be solved by the agent, using the existent problem solving resources. The cooperative medical diagnosis problems solving by the diagnosis system is partially based on the blackboard-based problem solving [18, 19]. The problem solving by the BMDS system is similar with the situations, when more physicians with different medical specializations plans a treatment to cure an illness that is in an advanced stage. Treatments known to be effective for the curing of the illness in a less advanced stage cannot be applied.

From the entire discussions one can comprehend and classify the medical agent-based IDSS research [20] into two categories, namely Clinical Management and Clinical Research. Clinical Management envelops all clinical systems that are designed to help the doctor with diagnosing and deciding on treatment for medical conditions. Clinical Research on the erstwhile envelopes systems that are used to research facts and connections in attempt to detect new trends and patterns; it covers systems for both diagnosing patients and treating them.

III. PREPROCESSING AND CLASSIFICATION

**3.1 Filter agent**

Feature selection, as a preprocessing step to machine learning, is effective in reducing dimensionality, removing irrelevant data, increasing learning accuracy, and improving result comprehensibility. In this work, we introduce a novel concept, predominant correlation, and propose a faster method which can identify relevant features as well as redundancy among relevant features without pair wise correlation analysis. The efficiency and effectiveness of our method is demonstrated through extensive comparisons with other methods using real-world data of high dimensionality.

## 3.2 Wrapper agent

In Wrapper based feature selection, the more states that are visited during the search phase of the algorithm the greater the likelihood of finding a feature subset that has a high internal accuracy while generalizing poorly. It removes the irrelevant attributes that are below the threshold value.

## 3.3 Classifier agent

The classifier agent uses the naïve Bayesian classification algorithm. Bayesian network algorithm is used to classify the collected attributes in to five classes (0-normal,1-starting,2-low,3-mild,4serious). The mutual information between each symptom and disease is computed based on information entropy theory. F and C are symptom and disease. f, c are events of F and C. $I(F,C)= P(f,c)logP(f,c)/P(f)P(c)$ Suppose I0 is the prior entropy of C. $I0 = (P(c=1)logp(c=1)+p(c=0)logp(c=0))$ The significance of each symptom is calculated by $S(F,C)=I(F,C)/I0$. All the symptoms are evaluated and ranked by significance index $S(F,C)$.

**Input Attributes Documentation**

1 id: patient identification number
2 ccf: social security number (I replaced this with a dummy value of 0)
3 age: age in years
4 sex: sex (1 = male; 0 = female)
5 painloc: chest pain location (1 = substernal; 0 = otherwise)
6 painexer (1 = provoked by exertion; 0 = otherwise)
7 relrest (1 = relieved after rest; 0 = otherwise)
8 pncaden (sum of 5, 6, and 7)
9 cp: chest pain type
  --Value 1: typical angina
  --Value 2: atypical angina
  --Value 3: non-anginal pain
  --Value 4: asymptomatic
10 trestbps: resting blood pressure (in mm Hg on admission to the hospital)
11 htn
12 chol: serum cholestoral in mg/dl
13 smoke: I believe this is 1 = yes; 0 = no (is or is not a smoker)
14 cigs (cigarettes per day)
15 years (number of years as a smoker)
16 fbs: (fasting blood sugar > 120 mg/dl) (1 = true; 0 = false)
17 dm (1 = history of diabetes; 0 = no such history)
18 famhist: family history of coronary artery disease (1 = yes; 0 = no)
19 restecg: resting electrocardiographic results
   --Value 0: normal
   --Value 1: having ST-T wave abnormality (T wave inversions and/or ST elevation or depression of > 0.05 mV)
   --Value 2: showing probable or definite left ventricular hypertrophy by Estes' criteria
20 ekgmo (month of exercise ECG reading)
21 ekgday(day of exercise ECG reading)
22 ekgyr (year of exercise ECG reading)
23 dig (digitalis used furing exercise ECG: 1 = yes; 0 = no)
24 prop (Beta blocker used during exercise ECG: 1 = yes; 0 = no)
25 nitr (nitrates used during exercise ECG: 1 = yes; 0 = no)
26 pro (calcium channel blocker used during exercise ECG: 1 = yes; 0 = no)
27 diuretic (diuretic used used during exercise ECG: 1 = yes; 0 = no)
28 proto: exercise protocol
    1 = Bruce
    2 = Kottus
    3 = McHenry
    4 = fast Balke
    5 = Balke
    6 = Noughton
    7 = bike 150 kpa min/min (Not sure if "kpa min/min" is what was written!)
    8 = bike 125 kpa min/min
    9 = bike 100 kpa min/min
    10 = bike 75 kpa min/min
    11 = bike 50 kpa min/min
    12 = arm ergometer
29 thaldur: duration of exercise test in minutes
30 thaltime: time when ST measure depression was noted
31 met: mets achieved
32 thalach: maximum heart rate achieved

33 thalrest: resting heart rate
34 tpeakbps: peak exercise blood pressure (first of 2 parts)
35 tpeakbpd: peak exercise blood pressure (second of 2 parts)
36 dummy
37 trestbpd: resting blood pressure
38 exang: exercise induced angina (1 = yes; 0 = no)
39 xhypo: (1 = yes; 0 = no)
40 oldpeak = ST depression induced by exercise relative to rest
41 slope: the slope of the peak exercise ST segment
    --Value 1: upsloping
    --Value 2: flat
    --Value 3: downsloping
42 rldv5: height at rest
43 rldv5e: height at peak exercise
44 ca: number of major vessels (0-3) colored by flourosopy
45 restckm: irrelevant
46 exerckm: irrelevant
47 restef: rest raidonuclid (sp?) ejection fraction
48 restwm: rest wall (sp?) motion abnormality
    0 = none
    1 = mild or moderate
    2 = moderate or severe
    3 = akinesis or dyskmem (sp?)
49 exeref: exercise radinalid (sp?) ejection fraction
50 exerwm: exercise wall (sp?) motion
51 thal: 3 = normal; 6 = fixed defect; 7 = reversable defect
52 thalsev: not used
53 thalpul: not used
54 earlobe: not used
55 cmo: month of cardiac cath (sp?) (perhaps "call")
56 cday: day of cardiac cath (sp?)
57 cyr: year of cardiac cath (sp?)
58 num: diagnosis of heart disease (angiographic disease status)
    --Value 0: < 50% diameter narrowing
    --Value 1: > 50% diameter narrowing
    (in any major vessel: attributes 59 through 68 are vessels)
59 lmt
60 ladprox
61 laddist
62 diag
63 cxmain
64 ramus
65 om1
66 om2
67 rcaprox
68 rcadist
69 lvx1: not used
70 lvx2: not used
71 lvx3: not used
72 lvx4: not used
73 lvf: not used
74 cathef: not used
75 junk: not used
76 name: last name of patient
    (I replaced this with the dummy string "name")

## IV. EXPERIMENTAL RESULTS

*Table: 1  Data Sets Used*

| S.No. | DATA SET NAME | NO. OF INSTANCES |
|---|---|---|
| 1 | CLEVELAND | 303 |
| 2 | HUNGARIAN | 294 |
| 3 | SWITZERLAND | 123 |
| 4 | LONG BEACH | 200 |

*Table: 2  Preprocessed Results-Cleveland Data Set*

| PID | FILTER | WRAPPER |
|---|---|---|
| 1 | 12 | 13 |
| 2 | 10 | 12 |
| 3 | 11 | 7 |
| 4 | 14 | 10 |
| 5 | 11 | 13 |
| 6 | 12 | 12 |
| 7 | 22 | 15 |
| 8 | 11 | 12 |
| 9 | 11 | 10 |
| 10 | 9 | 10 |

*Table: 3  Preprocessed Results-Hungarian Data Set*

| PID | FILTER | WRAPPER |
|---|---|---|
| 1254 | 10 | 12 |
| 1255 | 9 | 11 |
| 1256 | 8 | 11 |
| 1257 | 7 | 10 |

| PID | FILTER | WRAPPER |
|---|---|---|
| 1258 | 12 | 13 |
| 1259 | 11 | 10 |
| 1260 | 13 | 12 |
| 1261 | 11 | 10 |
| 1262 | 10 | 12 |
| 1263 | 13 | 13 |

*Table: 4 Preprocessed Results-Swiz Data Set*

| PID | FILTER | WRAPPER |
|---|---|---|
| 3001 | 11 | 6 |
| 3002 | 10 | 8 |
| 3003 | 11 | 9 |
| 3004 | 8 | 11 |
| 3005 | 9 | 10 |
| 3006 | 10 | 11 |
| 3007 | 10 | 11 |
| 3008 | 12 | 11 |
| 3009 | 11 | 11 |
| 3010 | 11 | 10 |

*Table: 5 Preprocessed Results-Longbeach Data Set*

| PID | FILTER | WRAPPER |
|---|---|---|
| 1 | 11 | 13 |
| 2 | 10 | 8 |
| 3 | 11 | 12 |
| 4 | 13 | 11 |
| 5 | 14 | 6 |
| 6 | 12 | 9 |
| 7 | 11 | 10 |
| 8 | 12 | 10 |
| 9 | 6 | 11 |
| 10 | 11 | 14 |

*Table: 6 Prior Probability of Symptoms*

| SYMPTOM | ABSENCE | STARTING | MILD | MODERATE | SERIOUS |
|---|---|---|---|---|---|
| PAINLOC | 0.3 | 0.4 | 0.6 | 0.7 | 0.9 |
| PAINEXER | 0.1 | 0.4 | 0.7 | 0.85 | 0.89 |
| RELREST | 0.2 | 0.34 | 0.4 | 0.7 | 0.8 |
| PNCADEN | 0.1 | 0.43 | 0.5 | 0.6 | 0.7 |
| CP | 0.3 | 0.2 | 0.3 | 0.4 | 0.8 |
| TRESTBPS | 0.3 | 0.4 | 0.45 | 0.5 | 0.76 |
| HTN | 0.2 | 0.2 | 0.4 | 0.5 | 0.7 |
| CHOL | 0.3 | 0.3 | 0.5 | 0.6 | 0.778 |
| SMOKE | 0.23 | 0.4 | 0.45 | 0.5 | 0.8 |
| CIGS | 0.14 | 0.4 | 0.4 | 0.5 | 0.9 |
| YEARS | 0.15 | 0.3 | 0.4 | 0.466 | 0.8 |
| FBS | 0.16 | 0.2 | 0.3 | 0.4 | 0.7 |
| DM | 0.3 | 0.3 | 0.4 | 0.5 | 0.888 |
| FAMHIST | 0.3 | 0.333 | 0.388 | 0.5 | 0.677 |
| RESTECG | 0.1 | 0.12 | 0.2 | 0.3 | 0.55 |
| EKGMO | 0.2 | 0.3 | 0.4 | 0.5 | 0.6 |
| EKGDAY | 0.3 | 0.3 | 0.4 | 0.488 | 0.7 |
| EKGYR | 0.2 | 0.3 | 0.395 | 0.4 | 0.9 |
| DIG | 0.23 | O.3 | 0.35 | 0.455 | 0.8 |

*TABLE: 7 CLASSIFICATION RESULTS*

| S.NO. | DATA SET | ABSENCE | STARTING | MILD | MODERATE | SERIOUS |
|---|---|---|---|---|---|---|
| 1 | CLEVELAND | 164 | 55 | 36 | 35 | 13 |
| 2 | HUNGARIAN | 188 | 37 | 26 | 28 | 15 |
| 3 | SWITZERLAND | 8 | 48 | 32 | 30 | 5 |
| 4 | LONG BEACH | 51 | 56 | 41 | 42 | 10 |

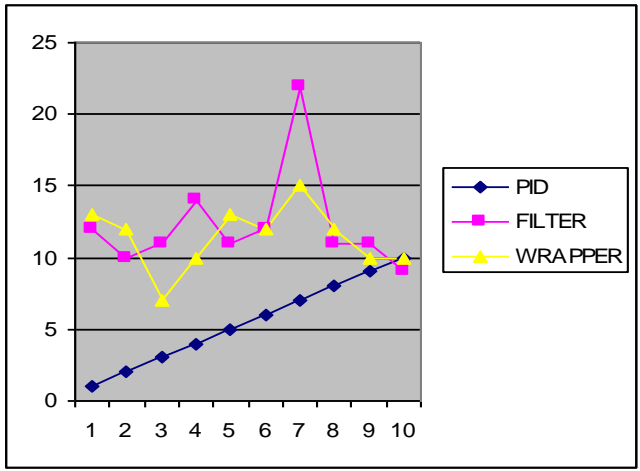

*Figure: 1    Preprocessed Chart for Cleveland Data Set*

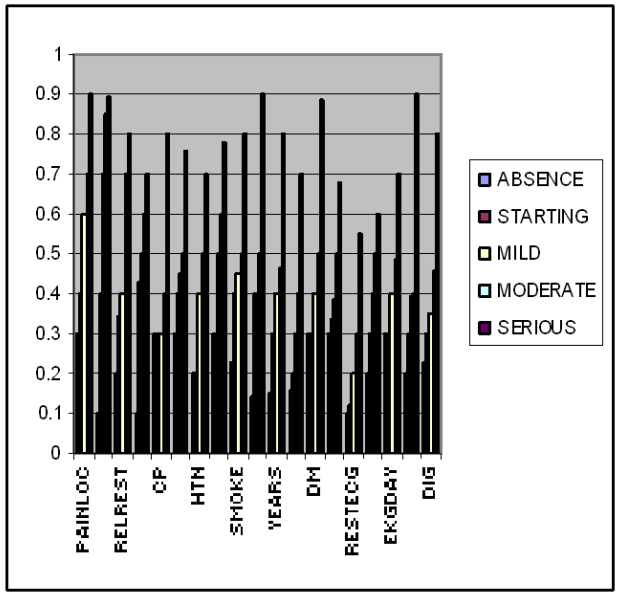

*Figure: 2    Classification Chart for all Data Sets*

## V. CONCLUSIONS AND FUTURE WORK

In the current work, input symptom values are preprocessed using filter and wrapper approach. Retained symptom values are used to classify cardiac patients in to five classes viz. Normal, Starting, Mild, Moderate and Serious. Classified values determine severity of the cardiac disease. Future work represents Heuristic seed selection to increase classification accuracy over the supervised naïve Bayesian classification. In this approach five seeds are selected for five classes. Future work represents diagnosis of cardiac patients using the co-operative elaboration algorithm. It uses contract net protocol that allows autonomy and sharing among the agents to solve all cardiac problems. Diagnosis results will be accurate in all conditions and then intelligent based decision support system is used to retrieve the patient information from the database.

## VI.    ACKNOWLEDGEMENT

Primarily, my gratitude goes to Dr. D.Manjula, A.P, DCSE, Anna University Chennai, who guided me to carry out this research with interest and involvement. With profound reverence and high regards, I thank Dr. P. Narayanasamy, Professor and HOD, DISE, Anna University Chennai and Dr.R.Krishnamoorthy, Professor and Dean, Anna University Trichy, for the motivation provided by them during this research work which enabled me to complete this work successfully.

## AUTHORS PROFILE

K. Murugesan**,** is pursuing his Ph.D in Computer Science from Anna University and also working for Anna University as a Superintendent. His research interests include Data Mining, Multi agent based systems etc.,

D.Manjula is working for Anna University as an Assistant Professor. Her research interests include Natural Language Processing, Text Mining, Artificial Intelligence, Databases and Parallel Computing.